\documentclass[11pt]{article}
\usepackage{amsmath,amssymb,url}

\begin{document}

\begin{titlepage}
\begin{flushright}
    November 2023
\end{flushright}
\begin{center}
  \vspace{3cm}
  {\bf \Large Reconstruction of Type II Supergravities \\[0.2cm] via $O(d) \times O(d)$ Duality Invariants}
  \\  \vspace{2cm}
  Yoshifumi Hyakutake and Kiyoto Maeyama
  \\ \vspace{1cm}
  {\it College of Science, Ibaraki University, \\
   Bunkyo 2-1-1, Mito, Ibaraki 310-8512, Japan}
\end{center}

\vspace{2cm}
\begin{abstract}
  We reconstruct type II supergravities by using building blocks of $O(d) \times O(d)$ invariants.
  These invariants are obtained by explicitly analyzing $O(d) \times O(d)$ transformations of 10 dimensional
  massless fields. Similar constructions are done by employing double field theory or generalized geometry,
  but we completed the reconstruction within the framework of the supergravities.
\end{abstract}
\end{titlepage}

\section{Introduction} \label{sec:Intro}

Dualities among superstring theories play important roles to reveal both perturbative and 
non-perturbative aspects of superstring theories. 
Especially, type IIA superstring theory is related to type IIB superstring theory by T-duality, 
which interchanges Kaluza-Klein modes (KK modes) and winding modes of a compactified circle 
direction\cite{Kikkawa:1984cp,Sakai:1985cs}.
In the low energy limit, massive modes in the type II superstring theories are decoupled, and the effective actions
are well described by corresponding type II supergravity theories\cite{Schwarz:1983qr,Huq:1983im}. 
The T-duality transformations of background massless fields are well-known as 
Buscher rule\cite{Buscher:1987sk,Buscher:1987qj}.

When the superstring theories are toroidally compactified on $T^d$, the duality transformation can be 
generalized to $O(d,d)$ duality\cite{Narain:1985jj,Narain:1986am}. 
Actually it is argued in ref. \cite{Meissner:1991zj} that, by assuming all fields depend only on a time coordinate,
NS-NS sector in the low energy effective action, which consist of a graviton, a dilaton and Kalb-Ramond
B field (B field), can be rewritten in manifestly $O(d,d)$ invariant expression. 
And $O(d,d)$ invariance of the NS-NS sector in general background was confirmed 
in refs.~\cite{Hassan:1991mq,Maharana:1992my}. 
Furthermore, it is also proven that $O(d,d)$ invariance can be extended 
to all orders in $\alpha'$ corrections to the low energy effective action\cite{Sen:1991zi}.

$O(d,d)$ transformation of R-R sector has been investigated in refs.\cite{Hull:1994ys}-\cite{Hassan:1999mm}.
One approach is to note that R-R potentials fill up a spinor representation of $SO(d,d)$ duality 
transformation\cite{Hull:1994ys,Witten:1995ex}. 
The spinor representation of R-R potentials combined with B field was 
explicitly constructed when the compactified space was $T^3$\cite{Brace:1998xz}, 
and general case of $T^d$ compactification was completed in ref.~\cite{Fukuma:1999jt}.  
Another approach was done by Hassan in refs.~\cite{Hassan:1995je}-\cite{Hassan:1999mm},
where the consistency of the duality transformation with local supersymmetry transformation is imposed.
In this approach, the $O(d,d)$ transformations of dilatinos and gravitinos are explicitly written
in terms of 10 dimensional forms, and those of R-R potentials are derived in a bispinor form.
In the type II superstring theories, the formulation of superspace which is compatible with T-duality
was discussed in ref.~\cite{Siegel:1993th}, and 
inclusion of R-R fields and an application to AdS background were investigated 
in refs.~\cite{Hatsuda:2014qqa,Hatsuda:2014aza}.
Generalization of the ref.~\cite{Hassan:1999mm} to non-abelian T-duality was done
in refs.~\cite{Sfetsos:2010uq,Lozano:2011kb}.

Although the type II supergravities possesses $O(d,d)$ duality invariance, forms of the action 
are not manifestly invariant in terms of 10 dimensional fields.
There are two formalism to improve this point.
First one is a double field theory, which treats internal coordinates of winding modes and KK modes 
simultaneously\cite{Siegel:1993xq,Hull:2009mi,Aldazabal:2013sca}. 
$O(d,d)$ transformation is realized as a rotation among these $2d$ coordinates
and fields are generalized to behave as tensors under this coordinate transformation.
$O(d,d)$ invariant forms of the type II supergravities are discussed in the framework of
the double field theory in refs.~\cite{Hohm:2011zr,Jeon:2011vx,Jeon:2012hp}.
Second one is a generalized geometry, which treats tangent and cotangent bundles
of compactified manifold on equal footing\cite{Hitchin:2003cxu,Gualtieri:2003dx,Hitchin:2010qz}.
Lie bracket of two vector fields are also modified to Courant bracket to incorporate
B field transformation with general coordinate one.
$O(d,d)$ invariant forms of the type II supergravities are discussed in the framework of
the generalized geometry in ref.~\cite{Coimbra:2011nw}.

The double field theory or the generalized geometry played important roles to reveal the $O(d,d)$
invariant structure, however, it is not so clear to derive such structure within the framework of the 
type II supergravities. 
In this paper, we revisit $O(d) \times O(d)$ subgroup of the duality transformation discussed
in the ref.~\cite{Hassan:1999mm} to construct $O(d) \times O(d)$ invariants within the framework
of the type II supergravities. 
We review that $O(d) \times O(d)$ transformations of NS-NS fields and fermionic fields are 
completely written in terms of 10 dimensional fields, and construct $O(d) \times O(d)$ 
invariants by evaluating these.
The actions of the type II supergravities are completely written by combinations of these building blocks,
which are consistent with ones obtained in refs.~\cite{Jeon:2012hp,Coimbra:2011nw}.

This paper is organized as follows.
In section \ref{sec:Review}, we review the $O(d)\times O(d)$ duality transformations of fields shown 
in the ref.~\cite{Hassan:1999mm}. 
Especially we show that these transformations can be written by using 
10 dimensional fields\footnote{We neglect R-R fields since these are already completed in the 
ref.~\cite{Hassan:1999mm}.}.  
In section \ref{sec:Invs}, we construct $O(d)\times O(d)$ duality invariants for NS-NS fields and
fermionic ones. We also check that these duality invariants in the background of fundamental strings
and wave solutions, or NS5-branes and KK monopoles.
In section \ref{sec:Boson}, we construct NS-NS bosonic terms in the type II supergravities 
by using the duality invariants.
In section \ref{sec:Fermion}, we construct fermionic bilinear terms in the type II supergravities 
by duality invariants. 
Section \ref{sec:Conclu} is devoted to conclusions and discussions.
In Appendix \ref{sec:app}, we review the actions of the type II supergravities for NS-NS sector and
fermionic bilinear terms.

\section{Brief Review of $O(d)\times O(d)$ Transformations} \label{sec:Review}

In this section, we briefly review $O(d) \times O(d)$ transformations of massless fields in the type II supergravities.
We denote the 10 dimensional spacetime indices as $K, L, M, N, \cdots$. 
Non compact spacetime directions are labeled by $\mu, \nu, \cdots$ and 
compact $d$ dimensions are done by $\alpha, \beta, \cdots$.  
On the other hand, local Lorentz  indices are denoted  as $A, B, C, D, \cdots$.
Non compact local Lorentz indices are labeled by $i, j, \cdots$ and those for compact $d$ dimensions
are done by $a, b, \cdots$.

NS-NS fields of the type II supergravities are the graviton $G_{MN}$, the Kalb-Ramon field $B_{MN}$ and
the dilaton $\Phi$. Corresponding fields with compact spatial indices $g_{\alpha\beta}$ and $B_{\alpha\beta}$ 
are gathered into
\begin{alignat}{3}
  \mathcal{H} = \begin{pmatrix} g^{-1} & - g^{-1} B \\ B g^{-1} & g - B g^{-1} B \end{pmatrix}, \label{eq:H}
\end{alignat}
and the $O(d,d)$ transformation $\mathcal{O}$ for massless NS-NS fields is defined by\cite{Meissner:1991zj}
\begin{alignat}{3}
  \mathcal{H}' = \mathcal{O}^T \mathcal{H} \mathcal{O}, \qquad
  \mathcal{O}^T \hat{\eta} \, \mathcal{O} = \hat{\eta}, \qquad 
  \hat{\eta} = \begin{pmatrix} 0 & {\bf 1}_d \\ {\bf 1}_d & 0 \end{pmatrix}. \label{eq:O(d,d)tr}
\end{alignat}
Dimensionally reduced dilaton field $\Phi - \frac{1}{4} \text{log}\, \text{det} g$ is invariant 
under the $O(d,d)$ transformation.
$O(d) \times O(d)$ subgroup is expressed as\cite{Sen:1991zi}
\begin{alignat}{3}
  \mathcal{O} = \frac{1}{2} 
  \begin{pmatrix} \mathcal{S+R} & \mathcal{S-R} \\ \mathcal{S-R} & \mathcal{S+R} \end{pmatrix}, \qquad
  \mathcal{S}^T \mathcal{S} = \mathcal{R}^T \mathcal{R} = {\bf 1}_d. \label{eq:O(d)O(d)tr}
\end{alignat}
The case of $\mathcal{S}=\mathcal{R}$ corresponds to a part of general linear coordinate transformation.

From the eq.~(\ref{eq:O(d,d)tr}), it is possible to extract duality transformations of dimensionally reduced fields.
And these are gathered into duality transformations of original 10 dimensional fields.
Below we summarize $O(d) \times O(d)$ transformations of fields in 10 dimensions\cite{Hassan:1994mq}.
By introducing $10 \times 10$ matrices as
\begin{alignat}{3} 
  &Q_\pm = \frac{1}{2} (S+R) \mp \frac{1}{2} (S-R) (G\mp B) , \label{eq:Qpm}
  \\
  &S = \begin{pmatrix} {\bf 1}_{10-d} & 0 \\ 0 & \mathcal{S} \end{pmatrix}, \qquad
  R = \begin{pmatrix} {\bf 1}_{10-d} & 0 \\ 0 & \mathcal{R} \end{pmatrix}, \notag
\end{alignat}
the 10 dimensional inverse metric transforms as
\begin{alignat}{3} 
    {G'}^{-1} &= Q_\pm G^{-1} Q^T_\pm. \label{eq:dualmet}
\end{alignat}
Since the duality invariant which includes the dilaton field is written by $\Phi - \frac{1}{4} \text{log}\, \text{det} G$,
the duality transformation of the dilaton field is given by
\begin{alignat}{3}
  \Phi' = \Phi - \frac{1}{2} \text{log}\, \text{det} Q_\pm. \label{eq:dilaton}
\end{alignat}
The $\pm$ sign originates from actions to the world-sheet left and right moving modes, respectively.
From the eq.~(\ref{eq:dualmet}), it is possible to define $O(d) \times O(d)$ transformation of the vielbein as
\begin{alignat}{3}
    E^{\prime M}_{(\pm) A} &= Q^M_{\pm N} E^N{}_A. \label{eq:dualviel} 
\end{alignat}
Notice that $E^{\prime M}_{(\pm) A}$ are related by local Lorenz transformation of
\begin{alignat}{3}
    E^{\prime M}_{(+) A} &= E^{\prime M}_{(-) B} \Lambda^B{}_A, \qquad
    \Lambda^B{}_A = E^B{}_M Q^{-1 M}_-{}_N Q^N_{+ K} E^K{}_A. \label{eq:LorTr}
\end{alignat}
Thus local Lorentz frame of the left moving sector is obtained by twisting that of the right moving sector 
by $\Lambda^A{}_B$. So invariants under local Lorentz transformation, 
which are constructed out of $E^{\prime M}_{(+) A}$,
can always be written in terms of $E^{\prime M}_{(-) A}$.

Since two kinds of vielbein can be used after the duality transformation, the 3-form field strength 
$H_{ABC} = E^M{}_A E^N{}_B E^K{}_C H_{MNK}$ also transforms in two ways as\cite{Hassan:1999mm}
\begin{alignat}{3}
  H'_{(\pm)ABC} &= E^{\prime M}_{(\pm) A} E^{\prime N}_{(\pm) B} E^{\prime K}_{(\pm) C} H'_{MNK} \notag
  \\
  &= H_{ABC} - 3 G_{KM} Q_\pm^{-1}{}^M{}_N (S-R)^{NL} 
  W^\pm_{L[BC} E^K{}_{A]}. \label{eq:dualtrH}
\end{alignat}
Here $W^\pm_M{}^A{}_B$ are connections defined by using torsionless spin connection $\Omega_M{}^A{}_B$ as
\begin{alignat}{3}
  W^{\pm  A}_{M \ \ B} = \Omega_M{}^A{}_B \mp \frac{1}{2} H_M{}^A{}_B, \label{eq:TorSpinCon}
\end{alignat}
and the duality transformations are calculated as
\begin{equation}
  W^{\prime \pm\ \ \ A}_{(\pm) M\ B} = {W^{\pm A}_N}_B {Q^{-1 N}_\mp}_M. \label{eq:DualTorSpinCon}
\end{equation}
Notice that $W^{\prime \pm\ \ \ A}_{(\pm) M\ B}$ are constructed out of $E^{\prime M}_{(\pm) A}$, respectively.
Similarly $\Gamma^{\pm K}{}_{MN}$ are connections defined by using affine connection $\Gamma^K{}_{MN}$ as
\begin{alignat}{3}
  \Gamma^{\pm K}{}_{MN} &= {\Gamma^K}_{MN} \pm \frac{1}{2} H^K{}_{MN}, \label{eq:TorCon}
\end{alignat}
and the duality transformations are derived as
\begin{equation}
  {\Gamma^{\pm K}}_{MN} = {Q^K_\pm}_{K'}{\Gamma^{\pm K'}}_{M'N'} Q^{-1 M'}_{\mp \ \ \ M} 
  Q^{-1 N'}_{\pm \ \ \ N} - \partial_M( Q^{K}_{\pm L} ) Q^{-1 L}_{\pm \ \ \ N}.
  \label{eq:DualTorCon}
\end{equation}
Since the vielbein is not used in the eq.~(\ref{eq:TorCon}), there are no $(\pm)$ subscription in the above.

Next let us summarize duality transformations of gravitinos $\Psi_{\pm M}$ and dilatinos $\lambda_{\pm}$. 
In ref.~\cite{Hassan:1999mm}, these transformations are derived so as to be consistent 
with the local supersymmetry (\ref{eq:susytr}).
It is easy to check this for the gravitino $\Psi_{- M}$, and the result is
\begin{alignat}{3}
  \Psi^\prime_{- M} &= \Psi_{-N} Q^{-1 N}_{+\ \ \ M}, \qquad
  \epsilon'_- = \epsilon_-. \label{eq:trGtino-}
\end{alignat}
To derive the above, we used $Q^N_{\pm M}\partial_N = \partial_M$.
This holds because the derivatives of fields with respect to the compact directions are zero.
For the gravitino $\Psi_{+ M}$, the duality transformation $\Psi'_{+ M}$ is defined by using $E^{\prime M}_{(-) A}$
and the susy transformation becomes
\begin{alignat}{3}
  \delta_+ \Psi^\prime_{+M} &= 2 \Big( \partial_M + \frac{1}{4} W^{\prime +}_{(-)MAB} \Gamma^{AB} \Big)
  \epsilon'_+ + \cdots \notag 
  \\
  &= 2 U_+ \Big(\partial_M + \frac{1}{4} W^{\prime +}_{(+)MAB} \Gamma^{AB} \Big) U_+^{-1} \epsilon'_+ 
  + \cdots, \label{eq:dlPsip}
\end{alignat}
where $\Gamma^A$ is a gamma matrix in 10 dimensions.
In the above we ignored R-R fields, and used local Lorentz transformation to change
$E^{\prime M}_{(-) A}$ to $E^{\prime M}_{(+) A}$. 
$U_+$ is a spinor representation of the local Lorentz transformation of $\Lambda^{-1}$, and satisfies 
$U_+ \Gamma^A U_+^{-1} = \Lambda^{-1 A}{}_B \Gamma^B$.
The eq.~(\ref{eq:dlPsip}) is compatible with the duality transformation if we define
\begin{alignat}{3}
  \Psi^\prime_{+ M} &= U_+ \Psi_{+ N} Q^{-1 N}_{- \ \ \ M}, \qquad 
  \epsilon'_+ = U_+ \epsilon_+. \label{eq:trGtino+}
\end{alignat}

Finally we consider $O(d)\times O(d)$ duality transformations of dilatinos.
As in the case of the gravitinos, the duality transformations are derived so as to be consistent 
with the local supersymmetry (\ref{eq:susytr}).
\begin{alignat}{3}
  \delta_- \lambda^\prime_- &= 2 \Big( \Gamma_{(-)}^{\prime M} \partial_M \Phi^\prime 
  + \frac{1}{12} \Gamma^{ABC} H^\prime_{(-)ABC} \Big) \epsilon'_- + \cdots \notag 
  \\
  &= 2 \Big\{ \Gamma^M \partial_M \Phi
  - \frac{1}{2} \Gamma^B E^A{}_M Q^{-1}_-{}^M{}_N (S-R)^{NL} W^-_{L AB} \notag 
  \\
  &\quad 
  + \frac{1}{12} \Gamma^{ABC} H_{ABC} - \frac{1}{4} \Gamma^{ABC} E_{CM} Q^{-1 M}_{-\ \ \ N}(S-R)^{NL} 
  W^-_{LAB} \Big\} \epsilon_- + \cdots \notag 
  \\
  &= 2 \Big( \Gamma^M \partial_M \Phi + \frac{1}{12}\Gamma^{ABC} H_{ABC} \Big) \epsilon_- 
  - \frac{1}{2} \Gamma^C \Gamma^{AB} E_{CM} Q^{-1 M}_{-\ \ \ N}(S-R)^{NL} W^-_{LAB} +\cdots \notag 
  \\
  &= \delta_- \big( \lambda_- - Q^{-1 M}_{-\ \ \ N}(S-R)^{NL} \Gamma_M \Psi_{-L} \big) . 
\end{alignat}
In the second equality, we used the eq.~(\ref{eq:dualtrH}) and employed the fifth line of the eq.~(\ref{eq:trPhi2}). 
Thus the duality transformation of the dilatino $\lambda_-$ is compatible with the local supersymmetry if we define 
\begin{alignat}{3}
  \lambda'_- &= \lambda_- - Q^{-1 M}_{-\ \ \ N}(S-R)^{NL} \Gamma_M \Psi_{-L}. \label{eq:trDtino-}
\end{alignat}
As in the case of the gravitino, the duality transformation $\lambda'_+$ is defined by using $E^{\prime M}_{(-) A}$.
By taking into account the local Lorentz transformation, we obtain
\begin{alignat}{3}
  \lambda_+^\prime &= U_+ \big( \lambda_+ + Q^{-1 M}_{+\ \ \ N}(S-R)^{NL} \Gamma_M \Psi_{+L} \big).    
  \label{eq:trDtino+} 
\end{alignat}

\section{$O(d)\times O(d)$ Duality Invariants} \label{sec:Invs}

In this section we construct $O(d)\times O(d)$ duality invariants.
In order to find these, let us prepare useful relations for $Q_\pm^M{}_N$. 
$Q_\pm$ are defined in the $10 \times 10$ matrix notation as (\ref{eq:Qpm}), 
and by noting $S^T S=R^T R=1$, we obtain
\begin{alignat}{3}
  Q_\pm (S-R)^T &= \frac{1}{2}(S+R)(S-R)^T \mp \frac{1}{2} (S-R) (G \mp B)(S-R)^T \notag
  \\
  &= - (S-R) \Big\{ \frac{1}{2}(S+R) \pm \frac{1}{2} (S-R) (G \pm B) \Big\}^T \notag
  \\
  &= - (S-R) Q_\mp^T. \label{eq:Qpm2}
\end{alignat}
It is often useful to express the above as follows.
\begin{alignat}{3}
  Q_\pm^{-1} (S-R) &= - \big\{ Q_\mp^{-1} (S-R) \big\}^T. \label{eq:Qpm3}
\end{alignat}
On the other hand, from the eq.~(\ref{eq:Qpm}), $Q_+$ is written by $Q_-$ as
\begin{alignat}{3}
  Q_+ = Q_- - (S-R) G. \label{eq:Qp}
\end{alignat}
By multiplying $Q_\mp^{-1}$ from the left, we find
\begin{alignat}{3}
  Q_\mp^{-1} (S-R) G = \pm (1 - Q_\mp^{-1} Q_\pm). \label{eq:Qp2}
\end{alignat}
Combining the eqs.~(\ref{eq:Qpm3}) and (\ref{eq:Qp2}), we obtain a useful relation
\begin{alignat}{3}
  G Q_\pm^{-1} (S-R) &= - \big\{ Q_\mp^{-1} (S-R) G \big\}^T \notag
  \\
  &= \mp 1 \pm  \big\{ Q_\mp^{-1} Q_\pm \big\}^T. \label{eq:Qpm4}
\end{alignat}
This relation is often used to construct $O(d)\times O(d)$ duality invariants.

\subsection{Duality Invariants $S^\pm_{ABC}$ and $T^\pm_A$ for NS-NS Bosonic Fields}

Now we construct duality invariants for NS-NS bosonic fields. 
$O(d)\times O(d)$ transformation of $H_{ABC}$ in 10 dimensions is evaluated as follows.
\begin{alignat}{3}
  H'_{(\pm)ABC} &= H_{ABC} - 3 G_{KM} Q_\pm^{-1}{}^M{}_N (S-R)^{NL} 
  W^\pm_{L[BC} E^K{}_{A]} \notag
  \\
  &= H_{ABC} \pm 3 W^\pm_{[ABC]} \mp 3 Q^{-1}_\mp{}^L{}_N Q_\pm^N{}_K 
  W^\pm_{L[BC} E^K{}_{A]} \notag
  \\
  &= H_{ABC} \pm 3 W^\pm_{[ABC]} \mp 3 W^{\prime\pm}_{(\pm)[ABC]}, \label{eq:trH}
\end{alignat}
where $W^\pm_{ABC} = E^M{}_A W^\pm_{MBC}$ and 
$W^{\prime\pm}_{ABC} = E^{\prime M}_{(\pm)}{}_A W^{\prime\pm}_{(\pm)MBC}$.
In the second line, we used the eq.~(\ref{eq:Qpm4}).
Thus we find $O(d)\times O(d)$ duality invariant of the form
\begin{alignat}{3}
  S^\pm_{ABC} &\equiv H_{ABC} \pm 3 W^\pm_{[ABC]}
  = -\frac{1}{2}H_{ABC} \pm 3 \Omega_{[ABC]}. \label{eq:invS}
\end{alignat}
Note that these do not behave like tensors under general coordinate transformation
and $E^{\prime M}_{(+)}{}_A$ is used for the $+$ mode of the dual theory.
This means that $S^{\prime \pm}_{(\pm)ABC} = S^\pm_{ABC}$.

$O(d)\times O(d)$ transformation of the dilaton is given by the eq.~(\ref{eq:dilaton}),
and the derivative of that equation is evaluated as
\begin{alignat}{3}
  \partial_\mu \Phi' - \partial_\mu \Phi &= - \frac{1}{2} \partial_\mu \log \det Q_\pm \notag
  \\
  &= - \frac{1}{2} Q^{-1}_\pm{}^\alpha{}_\beta \partial_\mu Q_\pm{}^\beta{}_\alpha \notag
  \\
  &= \pm \frac{1}{4} Q^{-1}_\pm{}^\alpha{}_\beta (S-R)^{\beta\gamma} 
  \partial_\mu (G\mp B)_{\gamma\alpha} \notag
  \\
  &= \pm \frac{1}{2} e^a{}_\alpha Q^{-1}_\pm{}^\alpha{}_\beta (S-R)^{\beta\gamma} 
  W^\pm_{\gamma ai} e^i{}_\mu \notag
  \\
  &= \pm \frac{1}{2} E^A{}_M Q^{-1}_\pm{}^M{}_N (S-R)^{NL} W^\pm_{L AB} E^B{}_\mu \notag
  \\
  &= - \frac{1}{2} E^{MA} W^\pm_{M AB} E^B{}_\mu 
  + \frac{1}{2} E^{MA} Q^{-1}_\mp{}^L{}_N Q_\pm^N{}_M W^\pm_{L AB} E^B{}_\mu \notag
  \\
  &= - \frac{1}{2} E^{MA} W^\pm_{M AB} E^B{}_\mu 
  + \frac{1}{2} E_{(\pm)}^{\prime MA} W^{\prime \pm}_{(\pm)M AB} E_{(\pm)}^{\prime B}{}_\mu. \label{eq:trPhi2}
\end{alignat}
The eq.~(\ref{eq:Qpm4}) is used in the 6th line, and 
$E'{}_{(\pm)}^i{}_\mu = E_{(\pm)}^i{}_\mu$ is used in the last line. 
Since $\partial_\alpha \Phi = 0$ and $E^{MA}W^\pm_{MAB} E^B{}_\alpha=0$, we find
$O(d)\times O(d)$ invariant of the form
\begin{alignat}{3}
  T^\pm_N \equiv \partial_N \Phi - \frac{1}{2} W^{\pm A}{}_{AN}
  = \partial_N \Phi - \frac{1}{2} \Omega^{A}{}_{AN}, \label{eq:invT}
\end{alignat}
where $W^{\pm A}{}_{AN} = E^{MA} W^\pm_{M AB} E^B{}_N$.
Notice that in the dual theory $E^{\prime M}_{(+)}{}_A$ is used for the $+$ mode,
so we obtained $T^{\prime \pm}_{(\pm)N} = T^\pm_N$.
Since $T^\pm_M Q_\pm^M{}_N = T^\pm_N$ holds,
$T^\pm_A = E^M{}_A T^\pm_M$ is also $O(d)\times O(d)$ invariant.
Invariants which are similar to $S^\pm_{ABC}$ and $T^\pm_A$ are also constructed 
in the flux formulation of the double field theory\cite{Geissbuhler:2013uka}.

\subsection{Duality Invariants $\Theta_\pm$ for Fermionic Fields}

$O(d)\times O(d)$ transformations of the dilatinos in 10 dimensions are given by
\begin{alignat}{3}
  \lambda'_\pm &= U_\pm \big\{ \lambda_\pm \pm G_{KM} Q_\pm^{-1 M}{}_N (S-R)^{NL} E^K{}_A 
  \Gamma^A \Psi_{\pm L} \big\} \notag
  \\
  &= U_\pm \big\{ \lambda_\pm - E^M{}_A \Gamma^A \Psi_{\pm M} 
  + Q^{-1}_\mp{}^L{}_N Q_\pm^N{}_M E^M{}_A 
  \Gamma^A \Psi_{\pm L} \big\} \notag
  \\
  &= U_\pm \big\{ \lambda_\pm - E^M{}_A \Gamma^A \Psi_{\pm M} 
  + E^{\prime M}_{(\pm)}{}_A \Gamma^A U_\pm^{-1} \Psi'_{\pm M} \big\} \notag
  \\
  &= U_\pm \big\{ \lambda_\pm - E^M{}_A \Gamma^A \Psi_{\pm M} \big\}
  + E^{\prime M}_{(-)}{}_A \Gamma^A \Psi'_{\pm M}, \label{eq:trdil}
\end{alignat}
where $U_-=1$ and $U_+$ is a spinor representation of local Lorentz transformation 
whose corresponding vector representation is given by 
$\Lambda^A{}_B = E^A{}_M Q_-^{-1 M}{}_N Q_+^N{}_L E^L{}_B$.
We used the eq.~(\ref{eq:Qpm4}) in the second line, and 
$U_+ \Gamma^A U_+^{-1} = \Lambda^{-1 A}{}_B \Gamma^B$ in the 4th line.
Thus we find duality invariants up to local Lorentz transformation $U_\pm$.
\begin{alignat}{3}
  \Theta_\pm &= \lambda_\pm - E^M{}_A \Gamma^A \Psi_{\pm M}. \label{eq:invTheta}
\end{alignat}
Notice that the dual theory is written by $E^{\prime M}_{(-)}{}_A$ for $\Theta_\pm$.
This means $\Theta'_{(-)\pm} = U_\pm \Theta_\pm$, 
which is different from $S^\pm_{ABC}$ and $T^\pm_A$.
Similar expressions to the above are also obtained in the framework of the double field 
theory\cite{Jeon:2012hp} or generalized geometry\cite{Coimbra:2011nw}.

\subsection{Check of Duality Invariants for Classical Solutions}

Since we have constructed $O(d) \times O(d)$ invariants, let us evaluate these values
for classical solutions which exchange under T-duality.

First we consider classical solutions of fundamental strings and waves.
The solution of the fundamental strings is given by
\begin{alignat}{3}
  ds^2 &= - h_1^{-1} (dX^0)^2 + \sum_{i=1}^8 (dX^i)^2 + h_1^{-1} (dX^9)^2, \label{eq:F1}
  \\
  e^\Phi &= h_1^{-\frac{1}{2}}, \quad B_{09} = - 1 + h_1^{-1}, \quad
  h_1 = 1 + \frac{c_1}{r^6}, \notag  
\end{alignat}
where $r^2 = \sum_{i=1}^8 (X^i)^2$.
And nontrivial components of $O(d) \times O(d)$ invariants for this solution are evaluated as
\begin{alignat}{3}
  S^\pm_{\hat{i}\hat{0}\hat{9}} = \frac{1}{2} h_1^{-1} \partial_i h_1. \label{eq:SofF1}
\end{alignat}
On the other hand, the dual solution of the wave along $X^9$ direction is given by
\begin{alignat}{3}
  ds^2 &= - h_\text{w}^{-1} (dX^0)^2 + \sum_{i=1}^8 (dX^i)^2 
  + h_\text{w} \big( dX^9 - (1-h_\text{w}^{-1}) dX^0 \big)^2, \quad
  h_\text{w} = 1 + \frac{c_\text{w}}{r^6}.  \label{eq:wave}
\end{alignat}
And nontrivial components of $O(d) \times O(d)$ invariants for this solution are calculated as
\begin{alignat}{3}
  S^{\prime \pm}_{(-)\hat{i}\hat{0}\hat{9}} 
  = \mp \frac{1}{2} h_\text{w}^{-1} \partial_i h_\text{w}. \label{eq:Sofwave}
\end{alignat}
Here we used hats for local Lorentz indices. 
If we set $h_1 = h_\text{w}$, we obtain $S^\pm_{ABC} =  S^{\prime \pm}_{(\pm)ABC}$ 
because $E^{\prime M}_{(-)A} = E^{\prime M}_{(+)A}$ except for
$E^{\prime 9}_{(-)\hat{9}} = - E^{\prime 9}_{(+)\hat{9}}$.

Second we consider classical solutions of smeared NS5-branes and KK monopoles.
The solution of the NS5-branes smeared along $X^9$ direction is given by
\begin{alignat}{3}
  ds^2 &= - (dX^0)^2 + (dX^1)^2 + \cdots + (dX^5)^2 
  + h_5 \sum_{i=6,7,8} (dX^i)^2 + h_5 (dX^9)^2, \label{eq:NS5}
  \\
  e^\Phi &= h_5^{\frac{1}{2}}, \quad H_{ij9} = \epsilon_{ijk} \partial^k h_5, \quad
  h_5 = 1 + \frac{c_5}{r^6}, \notag  
\end{alignat}
where $r^2 = \sum_{i=6,7,8} (X^i)^2$.
And nontrivial components of $O(d) \times O(d)$ invariants for this solution are evaluated as
\begin{alignat}{3}
  S^\pm_{\hat{i}\hat{j}\hat{9}} &= - \frac{1}{2} h_5^{-\frac{3}{2}} 
  \epsilon_{ijk} \partial^k h_5, \quad
  T_{\hat{i}} &= - \frac{1}{4} h_5^{-\frac{3}{2}} \partial_i h_5. \label{eq:STofNS5}
\end{alignat}
On the other hand, the dual solution of the KK monopoles is given by
\begin{alignat}{3}
  ds^2 &=- (dX^0)^2 + (dX^1)^2 + \cdots + (dX^5)^2
  + h_\text{m} \sum_{i=6,7,8} (dX^i)^2 
  + h_\text{m}^{-1} \big( dX^9 - A_i dX^i \big)^2,  \label{eq:KKm}
  \\
  F_{ij} &= \partial_i A_j - \partial_j A_i = -\epsilon_{ijk} \partial^k h_\text{m}, \quad 
  h_\text{m} = 1 + \frac{c_\text{m}}{r^6}. \notag
\end{alignat}
And nontrivial components of $O(d) \times O(d)$ invariants for this solution become
\begin{alignat}{3}
  S^{\prime \pm}_{(-)\hat{i}\hat{j}\hat{9}} &= \pm \frac{1}{2} h_\text{m}^{-\frac{3}{2}}
  \epsilon_{ijk} \partial^k h_\text{m}, \quad
  T'_{(-)\hat{i}} = - \frac{1}{4} h_\text{m}^{-\frac{3}{2}} \partial_i h_\text{m}.
  \label{eq:STofKKm}
\end{alignat}
Here we used hats for local Lorentz indices.
If we set $h_5 = h_\text{m}$, we obtain $S^\pm_{ABC} =  S^{\prime \pm}_{(\pm)ABC}$ 
because $E^{\prime M}_{(-)A} = E^{\prime M}_{(+)A}$ except for
$E^{\prime 9}_{(-)\hat{9}} = - E^{\prime 9}_{(+)\hat{9}}$.

\section{Construction of NS-NS Bosonic Terms in Type II Supergravity via Duality Invariants} \label{sec:Boson}

Let us construct NS-NS bosonic terms in the type II supergravities by using duality invariants.
Building blocks are $S^\pm_{ABC}$, $T_A$ and $W^\pm_{MAB}$. 
The action consists of two derivative terms, so candidates are $S^\pm_{ABC}S^{\pm ABC}$,
$T_A T^A$ and $G^{MN} W^\pm_{MAB} W^\pm_{N}{}^{AB}=W^{\pm ABC} W^\pm_{ABC}$ 
multiplied by $E e^{-2\Phi}$.

First we evaluate $E e^{-2\Phi} S^\pm_{ABC}S^{\pm ABC}$.
\begin{alignat}{3}
  &E e^{-2\Phi} S^\pm_{ABC}S^{\pm ABC} \notag
  \\
  &= E e^{-2\Phi} \big\{ H_{ABC}H^{ABC} \pm 6 H^{ABC} W^\pm_{ABC} 
  + 9  W^{\pm ABC} W^\pm_{[ABC]} \big\} \notag
  \\
  &= E e^{-2\Phi} \big\{ H_{ABC}H^{ABC} \pm 6 H^{ABC} W^\pm_{ABC} 
  + 3  W^{\pm ABC} W^\pm_{ABC} -6  W^{\pm ABC} W^\pm_{BAC} \big\}. \label{eq:S2}
\end{alignat}
Next we calculate $E e^{-2\Phi} T^A T_A$.
\begin{alignat}{3}
  &4 E e^{-2\Phi} T^A T_A \notag
  \\
  &= E e^{-2\Phi} G^{MN} 
  \big( 2 \partial_M \Phi - E^{KA} W^\pm_{KAB} E^B{}_M \big)
  \big( 2 \partial_N \Phi - E^{LC} W^\pm_{LCD} E^D{}_N \big) \notag
  \\
  &= E e^{-2\Phi} \big\{ 4 \partial^A \Phi \partial_A \Phi
  + W^{\pm A}{}_{AC} W^{\pm B}{}_B{}^C \big\} 
  - 2 E \partial_M (e^{-2\Phi}) E^{MA} E^{NB} W^\pm_{NAB} \notag
  \\
  &= E e^{-2\Phi} \big\{ 4 \partial^A \Phi \partial_A \Phi 
  + W^{\pm A}{}_{AC} W^{\pm B}{}_B{}^C
  + 2 E^{MA} E^{NB} \partial_M W^\pm_{NAB} \notag
  \\&\qquad\qquad
  + 2 E^K{}_C (\partial_M E^C{}_K) E^{MA} E^{NB} W^\pm_{NAB} 
  + 2 (\partial_M E^{MA}) E^{NB} W^\pm_{NAB} \notag
  \\&\qquad\qquad
  + 2 E^{MA} (\partial_M E^{NB}) W^\pm_{NAB} \big\} 
  - 2 \partial_M \big( E e^{-2\Phi} E^{MA} E^{NB} W^\pm_{NAB} \big) \notag
  \\
  &= E e^{-2\Phi} \big\{ 4 \partial^A \Phi \partial_A \Phi 
  + W^{\pm A}{}_{AC} W^{\pm B}{}_B{}^C \notag
  \\&\qquad\qquad
  + E^{MA} E^{NB} (R^\pm_{MNAB} - W^\pm_{MA}{}^C W^\pm_{NCB}
  + W^\pm_{NA}{}^C W^\pm_{MCB}) \notag
  \\&\qquad\qquad
  - 2 \Omega^A{}_A{}^C W^{\pm B}{}_{BC} 
  +2 \Omega^{ACB} W^\pm_{CAB} \big\} 
  + 2 \partial_M \big( E e^{-2\Phi} W^{\pm A}{}_A{}^M \big) \notag
  \\
  &= E e^{-2\Phi} \big\{ 4 \partial^A \Phi \partial_A \Phi 
  + R^\pm \mp H^{ABC} W^\pm_{ABC} + W^{\pm ABC} W^\pm_{BAC} \big\} \label{eq:T2}
  \\
  &\quad\,
  + 2 \partial_M \big( E e^{-2\Phi} W^{\pm A}{}_A{}^M \big), \notag
\end{alignat}
where
\begin{alignat}{3}
  R^\pm_{ABMN} \equiv \partial_M W^\pm_{NAB} - \partial_N W^\pm_{MAB}
  + W^\pm_{MA}{}^C W^\pm_{NCB} - W^\pm_{NA}{}^C W^\pm_{MCB}. \label{eq:Rpm}
\end{alignat}
In the 4th equality, we used
\begin{alignat}{3}
  &\partial_M E^M{}_A + E^K{}_C (\partial_M E^C{}_K) E^M{}_A = E^{NB} \Omega_{NBA} 
  = \Omega^B{}_{BA}, \notag
  \\
  &E^{MA} (\partial_M E^{NB}) W^\pm_{NAB} = - E^{MA} E^{NB} (\partial_M E^C{}_N) W^\pm_{CAB}
  = \Omega^{ACB} W^\pm_{CAB}. \label{eq:rel}
\end{alignat}
Now we require invariance under general coordinate transformation. 
This means that $HW^\pm$ and $W^{\pm 2}$ terms should be removed by combining 
$E e^{-2\Phi} S^\pm_{ABC}S^{\pm ABC}$, $E e^{-2\Phi}T_A T^A$ and 
$E e^{-2\Phi} W^{\pm ABC} W^\pm_{ABC}$. This uniquely constrains the form of the combination
up to overall factor and the result becomes
\begin{alignat}{3}
  &E e^{-2\Phi} \Big( \frac{1}{6} S^{ABC} S_{ABC} + 4T^A T_A 
  - \frac{1}{2} W^{\pm ABC} W^\pm_{ABC} \Big) \notag
  \\
  &= E e^{-2\Phi} \Big( 4 \partial^A \Phi \partial_A \Phi 
  + R^\pm + \frac{1}{6} H^{ABC} H_{ABC} \Big)
  + 2 \partial_M \big( E e^{-2\Phi} W^{\pm A}{}_A{}^M \big) \notag
  \\
  &= E e^{-2\Phi} \Big( 4 \partial^A \Phi \partial_A \Phi 
  + R - \frac{1}{2\cdot 3!} H^{ABC} H_{ABC} \Big)
  + 2 \partial_M \big( E e^{-2\Phi} W^{\pm A}{}_A{}^M \big). \label{eq:NSboson}
\end{alignat}
Thus we construct NS-NS bosonic terms of the type II supergravities via $O(d)\times O(d)$ 
duality invariants. The Lagrangian is $O(d,d)$ invariant since it behaves as a scalar 
under general coordinate transformation and invariant under a constant shift of B field.
Notice that the dual theory for + mode is written in terms of $E^{\prime M}_{(+)A}$,
but it is possible to use local Lorentz transformation $E^{\prime M}_{(+)A}=E^{\prime M}_{(-)B} \Lambda^B{}_A$
to write the + mode of the dual theory in terms of $E^{\prime M}_{(-)A}$.

\section{Construction of Fermionic Bilinear Terms in Type II Supergravities via Duality Invariants} \label{sec:Fermion}

Let us construct fermionic bilinear terms in the type II supergravities by using duality invariants.
First we consider bilinear terms of the dilatinos. Since the duality invariant forms of the dilatinos are
given by $\Theta_\pm$, we would like to construct duality invariants which partially contain
\begin{alignat}{3}
  \overline{\Theta}_\pm \Gamma^M \partial_M \Theta_\pm. \label{eq:Th2}
\end{alignat}
These are not duality invariants nor scalars under local Lorentz transformation.
In order to recover the latter covariance, we add the connection $S^\pm_{ABC}$ as follows.
\begin{alignat}{3}
  &\overline{\Theta}_\pm \Gamma^M \partial_M \Theta_\pm 
  \pm \frac{1}{12} \overline{\Theta}_\pm \Gamma^{ABC} S^\pm_{ABC} \Theta_\pm \notag
  \\
  &= \overline{\Theta}_\pm \Gamma^M \Big( \partial_M + \frac{1}{4} W^\pm_{MAB} \Gamma^{AB} \Big)
  \Theta_\pm 
  \pm \frac{1}{12} \overline{\Theta}_\pm \Gamma^{ABC} H_{ABC} \Theta_\pm \notag
  \\
  &= \overline{\Theta}_\pm \Gamma^M D_M \Theta_\pm 
  \mp \frac{1}{24} \overline{\Theta}_\pm \Gamma^{ABC} H_{ABC} \Theta_\pm . \label{eq:Th2inv}
\end{alignat}
Here we used $\overline{\Theta}_\pm \Gamma^A \Theta_\pm = 0$ for Majorana fermions,
and $D_M$ is a covariant derivative with respect to the connection of $\Omega_{MAB}$.
In this case, $D_M = \partial_M + \frac{1}{4} \Omega_{MAB}\Gamma^{AB}$.
Thus the terms in the eq.~(\ref{eq:Th2inv}) are scalars under local Lorentz transformation.
Furthermore, these are $O(d)\times O(d)$ duality invariant as we show below.
The dual theory is written by $E^{\prime M}_{(-)}{}_A$ for the vielbein, and the dual of the above is written as
\begin{alignat}{3}
  &\overline{\Theta'}_{(-)\pm} \Gamma^{\prime M}_{(-)} \partial_M \Theta'_{(-)\pm}
  \pm \frac{1}{12} \overline{\Theta'}_{(-)\pm} \Gamma^{ABC} S^{\prime \pm}_{(-)ABC} \Theta'_{(-)\pm} \notag
  \\
  &= \overline{\Theta'}_{(-)\pm} \Gamma^{\prime M}_{(-)} 
  \Big( \partial_M + \frac{1}{4} W^{\prime \pm}_{(-)MAB} \Gamma^{AB} \Big) \Theta'_{(-)\pm}
  \pm \frac{1}{12} \overline{\Theta'}_{(-)\pm} \Gamma^{ABC} H'_{(-)ABC} \Theta'_{(-)\pm} \notag
  \\
  &= \overline{\Theta}_{\pm} U_\pm^{-1} \Gamma^{\prime M}_{(-)} 
  \Big( \partial_M + \frac{1}{4} W^{\prime \pm}_{(-)MAB} \Gamma^{AB} \Big) U_\pm \Theta_{\pm}
  \pm \frac{1}{12} \overline{\Theta}_{\pm} U_\pm^{-1} \Gamma^{ABC} H'_{(-)ABC} U_\pm \Theta_{\pm} \notag
  \\
  &= \overline{\Theta}_{\pm}  \Gamma^{\prime M}_{(\pm)} 
  \Big( \partial_M + \frac{1}{4} W^{\prime \pm}_{(\pm)MAB} \Gamma^{AB} \Big) \Theta_{\pm}
  \pm \frac{1}{12} \overline{\Theta}_{\pm} \Gamma^{ABC} H'_{(\pm)ABC} \Theta_{\pm} \notag
  \\
  &=\overline{\Theta}_{\pm} \Gamma^{\prime M}_{(\pm)} \partial_M \Theta_{\pm}
  \pm \frac{1}{12} \overline{\Theta}_{\pm} \Gamma^{ABC} S^{\prime \pm}_{(\pm)ABC} \Theta_{\pm} \notag
  \\
  &= \overline{\Theta}_\pm \Gamma^M \partial_M \Theta_\pm 
  \pm \frac{1}{12} \overline{\Theta}_\pm \Gamma^{ABC} S^\pm_{ABC} \Theta_\pm. \label{eq:Th2chk}
\end{alignat}
In the 3rd equality, we used local Lorentz covariance for the $+$ mode, such as
$U_+^{-1} \Gamma^{\prime M}_{(-)} U_+ = \Gamma^{\prime M}_{(+)}$.
Thus the terms of the eq.~(\ref{eq:Th2inv}) are $O(d,d)$ invariant.

Next we consider two derivative terms which consists of $\Psi_{\pm M}$ and $\Theta_\pm$. 
The duality invariants should partially contain
\begin{alignat}{3}
  \overline{\Psi}_{\pm M} G^{MN} \partial_N \Theta_\pm. \label{eq:PsiTh}
\end{alignat}
These are not duality invariants nor scalars under local Lorentz transformation. 
In order to make scalars under local Lorentz transformation, 
we need to add the connection term to the above. 
\begin{alignat}{3}
  &\overline{\Psi}_{\pm M} G^{MN} \Big( \partial_N + \frac{1}{4} W^\pm_{NAB} \Gamma^{AB} \Big) 
  \Theta_\pm \notag
  \\
  &= \overline{\Psi}_{\pm}^M D_M \Theta_\pm \mp \frac{1}{8} \overline{\Psi}_{\pm}^M H_{MAB} 
  \Gamma^{AB} \Theta_\pm. \label{eq:PsiThinv}
\end{alignat}
Furthermore, these are duality invariants as we show below.
\begin{alignat}{3}
  &\overline{\Psi'}_{\pm M} G^{\prime MN}
  \Big( \partial_N + \frac{1}{4} W^{\prime \pm}_{(-)NAB} 
  \Gamma^{AB} \Big) \Theta'_{(-)\pm} \notag
  \\
  &= \overline{\Psi}_{\pm M} G^{MK} Q^N_{\mp K} U_\pm^{-1} \Big( \partial_N 
  + \frac{1}{4} W^{\prime \pm}_{(-)NAB} \Gamma^{AB} \Big) U_\pm \Theta_\pm \notag
  \\
  &= \overline{\Psi}_{\pm M} G^{MK} Q^N_{\mp K} \Big( \partial_N 
  + \frac{1}{4} W^{\prime \pm}_{(\pm)NAB} \Gamma^{AB} \Big) \Theta_\pm \notag
  \\
  &=\overline{\Psi}_{\pm M} G^{MN} \Big( \partial_N + \frac{1}{4} W^\pm_{NAB} 
  \Gamma^{AB} \Big) \Theta_\pm. \label{eq:PsiThchk}
\end{alignat}
Thus the terms of the eq.~(\ref{eq:PsiThinv}) are $O(d,d)$ invariant.

Finally let us investigate two derivative terms which are bilinear of Majorana gravitinos. 
These should partially contain following terms.
\begin{alignat}{3}
  \overline{\Psi}_{\pm L} G^{LN} \Gamma^M \partial_M \Psi_{\pm N}. \label{eq:Psi2}
\end{alignat}
These are not duality invariants nor scalars under local Lorentz transformation. 
In order to recover the latter covariance, we add connection terms $S^\pm_{ABC}$ and 
$\Gamma^{\mp K}{}_{MN}$ as follows.
\begin{alignat}{3}
  &\overline{\Psi}_{\pm L} G^{LN} \Gamma^M \partial_M \Psi_{\pm N}
  \pm \frac{1}{12} \overline{\Psi}_{\pm L} G^{LN} \Gamma^{ABC} S^\pm_{ABC} \Psi_{\pm N}
  - \overline{\Psi}_{\pm L} G^{LN} \Gamma^M \Gamma^{\mp K}{}_{MN} \Psi_{\pm K} \notag
  \\
  &= \overline{\Psi}_{\pm L} G^{LN} \Gamma^M \Big( \partial_M + \frac{1}{4} W^\pm_{MAB} 
  \Gamma^{AB} \Big) \Psi_{\pm N}
  \pm \frac{1}{12} \overline{\Psi}_{\pm L} G^{LN} \Gamma^{ABC} H_{ABC} \Psi_{\pm N} \notag
  \\&\quad\,
  - \overline{\Psi}_{\pm L} G^{LN} \Gamma^M \Gamma^{\mp K}{}_{MN} \Psi_{\pm K} \notag
  \\
  &= \overline{\Psi}_{\pm}^N \Gamma^M D_M \Psi_{\pm N}
  \mp \frac{1}{24} \overline{\Psi}_{\pm}^N \Gamma^{ABC} H_{ABC} \Psi_{\pm N} 
  \mp \frac{1}{2} \overline{\Psi}_{\pm N} \Gamma_M H^{NMK} \Psi_{\pm K}. \label{eq:Psi2inv}
\end{alignat}
Note that $D_M \Psi_{\pm N} = (\partial_M + \frac{1}{4} \Omega_{MAB}\Gamma^{AB}) \Psi_{\pm N}
- \Gamma^K{}_{MN} \Psi_{\pm K}$.
These are scalars under local Lorentz transformation.
The first two terms are similar to the eq.~(\ref{eq:Th2chk}), so the transformations under 
$O(d) \times O(d)$ are also similar. 
One difference is on the derivatives of $Q^{-1}_\mp$, which is written as
\begin{alignat}{3}
  &\overline{\Psi'}_{\pm L} G^{\prime LN} \Gamma^{\prime M}_{(-)} \partial_M \Psi'_{\pm N}
  \pm \frac{1}{12} \overline{\Psi'}_{\pm L} G^{\prime LN} 
  \Gamma^{ABC} S^{\prime\pm}_{(-)ABC} \Psi'_{\pm N} \notag
  \\
  &= \overline{\Psi}_{\pm L} G^{LN} \Gamma^M \partial_M \Psi_{\pm N}
  \pm \frac{1}{12} \overline{\Psi}_{\pm L} G^{LN} \Gamma^{ABC} S^\pm_{ABC} 
  \Psi_{\pm N} \notag
  \\&\quad\,
  + (\partial_M Q^{-1 N'}_{\mp}{}_N) Q^N_{\mp K} \overline{\Psi}_{\pm L} G^{LK} \Gamma^M \Psi_{\pm N'}.
  \label{eq:Psi2chk}
\end{alignat}
On the other hand, the $O(d) \times O(d)$ transformations of the connections 
$\Gamma^{\mp K}{}_{MN}$ are given by the eq.~(\ref{eq:DualTorCon}),
and the third term in the eq.~(\ref{eq:Psi2inv}) transforms as
\begin{alignat}{3}
  &- \overline{\Psi'}_{\pm L} G^{\prime LN} \Gamma^{\prime M}_{(-)} 
  \Gamma^{\prime\mp K}{}_{MN} \Psi'_{\pm K} \notag
  \\
  &= - \overline{\Psi}_{\pm L} G^{LN'} Q^N_{\mp N'} U_\pm^{-1} \Gamma^{\prime M}_{(-)} U_\pm 
  \Gamma^{\prime\mp K}{}_{MN} \Psi_{\pm K'} Q^{-1 K'}_{\mp}{}_K \notag
  \\
  &= - \overline{\Psi}_{\pm L} G^{LN'} \Gamma^{M'} Q^{-1 K'}_{\mp}{}_K 
  \Gamma^{\prime\mp K}{}_{MN} Q^M_{\pm M'} Q^N_{\mp N'} \Psi_{\pm K'} \notag
  \\
  &= - \overline{\Psi}_{\pm L} G^{LN} \Gamma^M \Gamma^{\mp K}{}_{MN} \Psi_{\pm K}
  + Q^{-1 K'}_{\mp}{}_K (\partial_M Q^K_{\mp N})  \overline{\Psi}_{\pm L} G^{LN} \Gamma^{M} \Psi_{\pm K'}.
  \label{eq:Psi2chk2}
\end{alignat}
In the second equality, we used 
$U_\pm^{-1} \Gamma^{\prime M}_{(-)} U_\pm = \Gamma^{\prime M}_{(\pm)} = Q^M_{\pm M'} \Gamma^{M'}$.
Thus we see that the last term in the eq.~(\ref{eq:Psi2chk}) is cancelled by the last term in the eq.~(\ref{eq:Psi2chk2}).
The combinations of the eq.~(\ref{eq:Psi2inv}) are $O(d,d)$ invariant.

So far we constructed $O(d,d)$ invariants of (\ref{eq:Th2inv}), (\ref{eq:PsiThinv}) and (\ref{eq:Psi2inv}).
Then up to overall factor the Lagrangian is expressed as
\begin{alignat}{3}
  &E e^{-2\Phi} \bigg[ \overline{\Theta}_\pm \Gamma^M D_M \Theta_\pm 
  \mp \frac{1}{24} \overline{\Theta}_\pm \Gamma^{ABC} H_{ABC} \Theta_\pm 
  + c_1 \Big\{ \overline{\Psi}_{\pm}^M D_M \Theta_\pm \mp \frac{1}{8} \overline{\Psi}_{\pm}^M H_{MAB} 
  \Gamma^{AB} \Theta_\pm \Big\} \notag
  \\
  &\qquad\quad
  + c_2 \Big\{ \overline{\Psi}_{\pm}^N \Gamma^M D_M \Psi_{\pm N}
  \mp \frac{1}{24} \overline{\Psi}_{\pm}^N \Gamma^{ABC} H_{ABC} \Psi_{\pm N} 
  \mp \frac{1}{2} \overline{\Psi}_{\pm N} \Gamma_M H^{NMK} \Psi_{\pm K} \Big\} \bigg] \notag
  \\
  &= E e^{-2\Phi} \bigg[ \overline{\lambda}_\pm \Gamma^M D_M \lambda_\pm 
  - \overline{\lambda}_\pm \Gamma^M D_M (\Gamma^A \Psi_{\pm A}) 
  + \overline{\Psi}_{\pm A} \Gamma^A \Gamma^M D_M \lambda_\pm
  - \overline{\Psi}_{\pm A} \Gamma^A \Gamma^M D_M (\Gamma^B \Psi_{\pm B}) \notag
  \\&\qquad\qquad
  \mp \frac{1}{24} \overline{\lambda}_\pm \Gamma^{ABC} H_{ABC} \lambda_\pm
  \mp \frac{1}{12} \overline{\Psi}_{\pm D} \Gamma^D \Gamma^{ABC} H_{ABC} \lambda_{\pm}
  \pm \frac{1}{24} \overline{\Psi}_{\pm D} \Gamma^D \Gamma^{ABC} \Gamma^E H_{ABC} \Psi_{\pm E} \notag
  \\&\qquad\qquad
  + c_1 \Big\{ \overline{\Psi}_{\pm}^M D_M \lambda_\pm - \overline{\Psi}_{\pm}^M D_M (\Gamma^A \Psi_{\pm A})
  \mp \frac{1}{8} \overline{\Psi}_{\pm}^M H_{MAB} \Gamma^{AB} \lambda_\pm
  \pm \frac{1}{8} \overline{\Psi}_{\pm}^M H_{MAB} \Gamma^{AB} \Gamma^C \Psi_{\pm C} \Big\} \notag
  \\&\qquad\qquad
  + c_2 \Big\{ \overline{\Psi}_{\pm}^N \Gamma^M D_M \Psi_{\pm N}
  \mp \frac{1}{24} \overline{\Psi}_{\pm}^N \Gamma^{ABC} H_{ABC} \Psi_{\pm N} 
  \mp \frac{1}{2} \overline{\Psi}_{\pm N} \Gamma_M H^{NMK} \Psi_{\pm K} \Big\} \bigg] \notag
  \\
  &= E e^{-2\Phi} \bigg[ \overline{\lambda}_\pm \Gamma^M D_M \lambda_\pm 
  - \overline{\lambda}_\pm \Gamma^M \Gamma^A D_M \Psi_{\pm A}
  - \overline{\Psi}_{\pm A} \Gamma^M \Gamma^A D_M \lambda_\pm
  + \overline{\Psi}_{\pm A} \Gamma^M \Gamma^{AB} D_M \Psi_{\pm B} \notag
  \\&\qquad\qquad
  \mp \frac{1}{24} \overline{\lambda}_\pm \Gamma^{ABC} H_{ABC} \lambda_\pm
  \mp \frac{1}{12} \overline{\Psi}_{\pm D} \Gamma^{DABC} H_{ABC} \lambda_{\pm}
  \pm \frac{1}{24} \overline{\Psi}_{\pm D} \Gamma^{DABCE} H_{ABC} \Psi_{\pm E} \notag
  \\&\qquad\qquad
  + (2+c_1) \overline{\Psi}_{\pm}^M D_M \lambda_\pm 
  - (2+c_1) \overline{\Psi}_{\pm}^M \Gamma^A D_M \Psi_{\pm A}
  + (1+c_2) \overline{\Psi}_{\pm}^A \Gamma^M D_M \Psi_{\pm A} \notag
  \\&\qquad\qquad
  \mp \frac{2+c_1}{8} \overline{\Psi}_{\pm}^M H_{MAB} \Gamma^{AB} \lambda_\pm
  \pm \frac{2+c_1}{8} \overline{\Psi}_{\pm}^M H_{MAB} \Gamma^{AB} 
  \Gamma^C \Psi_{\pm C} \notag
  \\&\qquad\qquad
  \mp \frac{1+c_2}{24} \overline{\Psi}_{\pm}^N \Gamma^{ABC} H_{ABC} \Psi_{\pm N} 
  \mp \frac{1+2c_2}{4} \overline{\Psi}_{\pm N} \Gamma_M H^{NMK} \Psi_{\pm K} \bigg] . \label{eq:2Fermipre}
\end{alignat}
In the last equality, if we choose $c_1=-2$ and $c_2=-1$, 
it is possible to express the derivative of the Majorana gravitinos as field strengths 
of $D_{[M} \Psi_{\pm N]}$ up to partial integral. Since this prescription is important to realize
local supersymmetry, we employ these values.
Then the $O(d,d)$ invariant action of the fermionic bilinear is uniquely determined as
\begin{alignat}{3}
  &E e^{-2\Phi} \bigg[ \overline{\Theta}_\pm \Gamma^M D_M \Theta_\pm 
  \mp \frac{1}{24} \overline{\Theta}_\pm \Gamma^{ABC} H_{ABC} \Theta_\pm 
  -2 \Big\{ \overline{\Psi}_{\pm}^M D_M \Theta_\pm \mp \frac{1}{8} \overline{\Psi}_{\pm}^M H_{MAB} 
  \Gamma^{AB} \Theta_\pm \Big\} \notag
  \\
  &\qquad\quad
  - \Big\{ \overline{\Psi}_{\pm}^N \Gamma^M D_M \Psi_{\pm N}
  \mp \frac{1}{24} \overline{\Psi}_{\pm}^N \Gamma^{ABC} H_{ABC} \Psi_{\pm N} 
  \mp \frac{1}{2} \overline{\Psi}_{\pm N} \Gamma_M H^{NMK} \Psi_{\pm K} \Big\} \bigg] \notag
  \\
  &= E e^{-2\Phi} \bigg[ \overline{\lambda}_\pm \Gamma^M D_M \lambda_\pm 
  - \overline{\lambda}_\pm \Gamma^M \Gamma^A D_M \Psi_{\pm A}
  - \overline{\Psi}_{\pm A} \Gamma^M \Gamma^A D_M \lambda_\pm
  + \overline{\Psi}_{\pm A} \Gamma^M \Gamma^{AB} D_M \Psi_{\pm B} \notag
  \\&\qquad\qquad
  \mp \frac{1}{24} \overline{\lambda}_\pm \Gamma^{ABC} H_{ABC} \lambda_\pm
  \mp \frac{1}{12} \overline{\Psi}_{\pm D} \Gamma^{DABC} H_{ABC} \lambda_{\pm}
  \pm \frac{1}{24} \overline{\Psi}_{\pm D} \Gamma^{DABCE} H_{ABC} \Psi_{\pm E} \notag
  \\&\qquad\qquad 
  \pm \frac{1}{4} \overline{\Psi}_{\pm N} \Gamma_M H^{NMK} \Psi_{\pm K} \bigg] . \label{eq:2Fermi}
\end{alignat}
Of course, a linear combination of these terms is consistent with the type II supergravities.
Thus we showed that fermionic bilinears without R-R fields can be written in terms of the 
duality invariants within the framework of the type II supergravities.
Invariant forms of fermionic bilinears with R-R fluxes are obtained in the framework of the double field 
theory\cite{Jeon:2012hp} or generalized geometry\cite{Coimbra:2011nw}.

\section{Conclusion and Discussion} \label{sec:Conclu}
In this paper, within the framework of the type II supergravities, we have constructed 
$O(d)\times O(d)$ duality invariants of the eqs.~\eqref{eq:invS} \eqref{eq:invT} and 
\eqref{eq:invTheta} by examining $O(d)\times O(d)$ transformations of 3-form H field, dilaton 
and dilatino. These invariants are checked in the background of fundamental strings and wave solutions,
or NS5-branes and KK monopoles.
By using these duality invariants, we  reconstructed the actions of type II supergravities 
in a manifestly $O(d)\times O(d)$ invariant form in section \ref{sec:Boson} and \ref{sec:Fermion}.
Since these actions are also invariant under linear $GL(d)$ transformation and shift of the B field,
these are exactly $O(d,d)$ invariant. 
As for the kinetic terms on R-R fields, $SO(d,d)$ invariant construction was already discussed
within the framework of the type II supergravities in the ref.~\cite{Hassan:1999mm}.

As we have checked the duality invariants in the background of strings and wave solutions,
or NS5-branes and KK monopoles, it is easy to apply to other nongeometric 
backgrounds\cite{deBoer:2010ud}-\cite{Fernandez-Melgarejo:2018yxq}.
It is interesting to see corrections to the nongeometric background which was studied from
the viewpoint of world-sheet instantons\cite{Kimura:2013zva}. 
It is also interesting to investigate $\beta$-twisted solutions of the double field theory\cite{Sakamoto:2017cpu}
by evaluating $O(d)\times O(d)$ invariants in this paper.

Since we have constructed $O(d) \times O(d)$ duality invariants within the framework of the
type II supergravities, it is natural to generalize these formulation to higher derivative corrections
in the type II superstring theories. However, this is not a simple task and
it is shown that higher derivative corrections in bosonic or heterotic string theory cannot be written
in terms of generalized metric\cite{Hohm:2016lge,Hohm:2016yvc}.
We should take into account total derivative terms and field redefinitions which consist of dimensionally reduced fields.
Constraint on $R^2$ terms via cosmological ansatz was investigated in ref.~\cite{Meissner:1996sa}, 
was executed via T-duality in refs.~\cite{Garousi:2019wgz}, 
and was done via $O(d,d)$ duality in ref.~\cite{Eloy:2020dko}.
In our formalism, the difficulty can be seen by duality transformation of the Riemann tensor (\ref{eq:Rpm}),
which is calculated as
\begin{alignat}{3}
  R^{\prime \pm}_{(\pm) ABCD} &= E^{\prime M}_{(\pm) C} E^{\prime N}_{(\pm) D} 
  R^{\prime \pm}_{(\pm) ABMN} \notag
  \\
  &= R^{\pm}_{ABCD} \pm 2 R^{\pm}_{ABN[C} X^N_{\mp D]}
  + 2 X^M_{\mp [C} X^N_{\mp D]} W^\pm_{MAE} W^\pm_N{}^E{}_B \notag
  \\&\quad\,
  \mp 2 W^\pm_{KAB} W^\pm_{[CD]E} X_\mp^{KE}
  + 2 W^\pm_{KAB} W^\pm_{LE[C} X^L_{\mp D]} X_\mp^{KE} \label{eq:trRiemann}
  \\&\quad\,
  + X_\mp^{KE} S_{ECD} W^\pm_{KAB}, \notag
\end{alignat}
where $ X_\mp^{KE}=Q_{\mp}^{-1 K}{}_L(S-R)^{LE}=-X_\pm^{EK}$.
If we consider $R^\pm_{ABCD} S^{\pm ABE} S^{\pm CD}{}_E$, which exists as a part of higher derivative terms
in bosonic string theory, the duality transformation of this term
contains $\pm 2 R^{\pm}_{ABN[C} X^N_{\mp D]}S^{\pm ABE} S^{\pm CD}{}_E$.
However, this cannot be cancelled by other terms even if we consider total derivatives and field redefinitions of 
10 dimensional fields.

Although we should decompose the eq.~(\ref{eq:trRiemann}) in terms of dimensionally reduced
fields, $S_{ABC}$, $T_A$ and $W^\pm_{\mu AB}$ are invariant under $O(d,d)$ transformations.
Thus we should only take care of $W^\pm_{\alpha AB}$. 
It is also useful to consult a frame formalism of the double field theory\cite{Marques:2015vua}.
If we find nice structure on total derivatives and field redefinitions
in terms of these fields, it will be possible to apply our $O(d)\times O(d)$ construction to higher derivative terms
such as $R^4$ terms\cite{Hohm:2019jgu}-\cite{Garousi:2022ghs}.

\section*{Acknowledgement}

The authors would like to thank Takanori Fujiwara and Makoto Sakaguchi.
YH would like to thank Tetsuji Kimura and Kentaroh Yoshida for useful conversations.
We would also like to thank Yuho Sakatani for valuable comments.
This work was partially supported by Japan Society for the Promotion of Science, 
Grant-in-Aid for Scientific Research (C) Grant Number 22K03613.

\appendix
\section{Review of Type II Supergravities}\label{sec:app}

Let us ignore R-R fields of the type II supergravities.
Then Lagrangian of the type IIA supergravity is given by\cite{Huq:1983im}
\begin{alignat}{3}
  \mathcal{L} &= E e^{-2\Phi} \bigg[ 4 \partial^A \Phi \partial_A \Phi 
  + R - \frac{1}{2\cdot 3!} H^{ABC} H_{ABC} \notag
  \\&\qquad\qquad
  + \overline{\lambda} \Gamma^M D_M \lambda
  - \overline{\lambda} \Gamma^M \Gamma^A D_M \Psi_A
  - \overline{\Psi}_A \Gamma^M \Gamma^A D_M \lambda
  + \overline{\Psi}_A \Gamma^M \Gamma^{AB} D_M \Psi_B \notag
  \\&\qquad\qquad
  - \frac{1}{24} \overline{\lambda} \Gamma^{ABC} H_{ABC} \Gamma_{11} \lambda
  - \frac{1}{12} \overline{\Psi}_D \Gamma^{DABC} H_{ABC} \Gamma_{11} \lambda
  - \frac{1}{24} \overline{\Psi}_D \Gamma^{DABCE} H_{ABC} \Gamma_{11} \Psi_E \notag
  \\&\qquad\qquad 
  - \frac{1}{4} \overline{\Psi}_N \Gamma_M H^{NMK} \Gamma_{11} \Psi_K \notag
  \\
  &= E e^{-2\Phi} \bigg[ 4 \partial^A \Phi \partial_A \Phi 
  + R - \frac{1}{2\cdot 3!} H^{ABC} H_{ABC} \notag
  \\&\qquad\qquad
  + \overline{\lambda}_+ \Gamma^M D_M \lambda_+
  - \overline{\lambda}_+ \Gamma^M \Gamma^A D_M \Psi_{+ A}
  - \overline{\Psi}_{+ A} \Gamma^M \Gamma^A D_M \lambda_+
  + \overline{\Psi}_{+ A} \Gamma^M \Gamma^{AB} D_M \Psi_{+ B} \notag
  \\&\qquad\qquad
  - \frac{1}{24} \overline{\lambda}_+ \Gamma^{ABC} H_{ABC} \lambda_+
  - \frac{1}{12} \overline{\Psi}_{+ D} \Gamma^{DABC} H_{ABC} \lambda_+
  + \frac{1}{24} \overline{\Psi}_{+ D} \Gamma^{DABCE} H_{ABC} \Psi_{+ E} \notag
  \\&\qquad\qquad 
  + \frac{1}{4} \overline{\Psi}_{+ N} \Gamma_M H^{NMK} \Psi_{+ K} \label{eq:typeIIA}
  \\&\qquad\qquad
  + \overline{\lambda}_- \Gamma^M D_M \lambda_- 
  - \overline{\lambda}_- \Gamma^M \Gamma^A D_M \Psi_{- A}
  - \overline{\Psi}_{- A} \Gamma^M \Gamma^A D_M \lambda_-
  + \overline{\Psi}_{- A} \Gamma^M \Gamma^{AB} D_M \Psi_{- B} \notag
  \\&\qquad\qquad
  + \frac{1}{24} \overline{\lambda}_- \Gamma^{ABC} H_{ABC} \lambda_-
  + \frac{1}{12} \overline{\Psi}_{- D} \Gamma^{DABC} H_{ABC} \lambda_-
  - \frac{1}{24} \overline{\Psi}_{- D} \Gamma^{DABCE} H_{ABC} \Psi_{- E} \notag
  \\&\qquad\qquad 
  - \frac{1}{4} \overline{\Psi}_{- N} \Gamma_M H^{NMK} \Psi_{- K} \bigg]. \notag
\end{alignat}
Here $\lambda$ and $\Psi_M$ are Majorana fermions and satisfy
\begin{alignat}{3}
  &\lambda = \lambda_+ + \lambda_-, \quad \Psi_M = \Psi_{+ M} + \Psi_{-M}, \notag
  \\
  &\Gamma_{11} \lambda_\pm = \pm \lambda_\pm, \quad \Gamma_{11} \Psi_{\pm M} = \mp \Psi_{\pm M}.
  \label{eq:gammaIIA}
\end{alignat}
We chose similar notations as in the ref.~\cite{Bergshoeff:2001pv,Imamura2007}.
Lagrangian of the type IIB supergravity takes a similar form as the eq.~(\ref{eq:typeIIA}), but $\pm$ modes of the
dilatinos or the gravitinos have the same chirality.
\begin{alignat}{3}
  &\Gamma_{11} \lambda_\pm = - \lambda_\pm, \quad \Gamma_{11} \Psi_{\pm M} = \Psi_{\pm M}.
  \label{eq:gammaIIB}
\end{alignat}

In the case of the type IIA supergravity, transformations of massless fields under local supersymmetry are given by
\begin{alignat}{3}
  \delta E^A{}_M &= \bar{\epsilon} \Gamma^A \Psi_M 
  = \bar{\epsilon}_+ \Gamma^A \Psi_{+ M} + \bar{\epsilon}_- \Gamma^A \Psi_{- M}, \notag
  \\
  \delta \Phi &= \frac{1}{2} \bar{\epsilon} \lambda
  = \frac{1}{2} \bar{\epsilon}_+ \lambda_- + \frac{1}{2} \bar{\epsilon}_- \lambda_+, \notag
  \\
  \delta B_{MN} &= 2 \bar{\epsilon} \Gamma_{11} \Gamma_{[M} \Psi_{N]}
  = 2 \bar{\epsilon}_+ \Gamma_{[M} \Psi_{+ N]} - 2 \bar{\epsilon}_- \Gamma_{[M} \Psi_{- N]}, \label{eq:susytr}
  \\
  \delta \lambda &= 2 (\partial_M \Phi) \Gamma^M \epsilon
  + \frac{1}{6} H_{ABC} \Gamma^{ABC} \Gamma_{11} \epsilon \notag
  \\
  &= 2 \Big\{ (\partial_M \Phi) \Gamma^M - \frac{1}{12} H_{ABC} \Gamma^{ABC} \Big\} \epsilon_+ 
  + 2 \Big\{ (\partial_M \Phi) \Gamma^M + \frac{1}{12} H_{ABC} \Gamma^{ABC} \Big\} \epsilon_-, \notag
  \\
  \delta \Psi_M &= 2 D_M \epsilon + \frac{1}{4} H_{MAB} \Gamma^{AB} \Gamma_{11} \epsilon \notag
  \\
  &= 2 \Big( \partial_M + \frac{1}{4} W^+_{MAB} \Gamma^{AB} \Big) \epsilon_+
  + 2 \Big( \partial_M + \frac{1}{4} W^-_{MAB} \Gamma^{AB} \Big) \epsilon_- . \notag
\end{alignat}
Again we ignored contributions of R-R fields.
$\epsilon$ is a Majorana fermion and satisfy
\begin{alignat}{3}
  &\epsilon = \epsilon_+ + \epsilon_-, \quad \Gamma_{11} \epsilon_\pm = \mp \epsilon_\pm. \label{eq:IIAep}
\end{alignat}
In the case of the type IIB supergravity, $\epsilon_\pm$ should satisfy
\begin{alignat}{3}
  &\Gamma_{11} \epsilon_\pm = \epsilon_\pm. \label{eq:IIBep}
\end{alignat}


\end{document}